\documentclass[prd,showpacs,twocolumn,
amsmath,amssymb,superscriptaddress,floatfix,nofootinbib]{revtex4-1}
\usepackage[utf8]{inputenc}
\usepackage[sort&compress]{natbib}
\usepackage[normalem]{ulem}
\usepackage{lipsum} 
\usepackage{bm}
\usepackage{times}
\usepackage{amssymb,amsbsy,amsmath,amsfonts}
\usepackage{graphicx}
\usepackage{float}
\usepackage{color}
\usepackage{morefloats}
\usepackage{rotating}
\usepackage{srcltx}
\usepackage{slashed}
\usepackage{subfigure}
\usepackage{multirow}
\usepackage{verbatim}
\usepackage{hyperref}
\usepackage{tabularx}
\usepackage{adjustbox}
\usepackage[dvipsnames]{xcolor}
\usepackage{nicefrac}

\usepackage{hyperref}
\hypersetup{
	colorlinks	=true,
	urlcolor	=blue,
	linkcolor	=blue,
	citecolor	=blue,
	pdftitle	={},
	pdfauthor	={Ru-You Zheng,Zhi-Wei Liu, Li-Sheng Geng},
	pdfsubject	={$\Lambda_c N$ correlation function}
	}

\begin{document}

\title{
$\Lambda_c N$ correlation functions with leading-order covariant chiral interactions
}

\author{Ru-You Zheng}
\affiliation{School of Physics, Beihang University, Beijing 102206, China}
\affiliation{Research Center for Nuclear Physics (RCNP), Ibaraki, Osaka 567-0047, Japan}

\author{Zhi-Wei Liu}
\email[Corresponding author: ]{liuzhw@buaa.edu.cn}
\affiliation{Institute for Advanced Study in Nuclear Energy \& Safety, College of Physics and Optoelectronic Engineering, Shenzhen University, Shenzhen 518060,
Guangdong, China}

\author{Li-Sheng Geng}
\email[Corresponding author: ]{lisheng.geng@buaa.edu.cn}
\affiliation{Sino-French Carbon Neutrality Research Center, \'{E}cole Centrale de P\'{e}kin/School
of General Engineering, Beihang University, Beijing 100191, China}
\affiliation{School of Physics, Beihang University, Beijing 102206, China}
\affiliation{Peng Huanwu Collaborative Center for Research and Education, Beihang University, Beijing 100191, China}
\affiliation{Southern Center for Nuclear-Science Theory (SCNT), Institute of Modern Physics, Chinese Academy of Sciences, Huizhou 516000, China}

\begin{abstract}
The $\Lambda_c p$ momentum correlation functions are investigated using $\Lambda_c N$ interactions derived within the covariant chiral effective field theory. Our analysis reveals that the interaction is weakly attractive in the spin-singlet ${}^1S_0$ channel. In contrast, the ${}^3S_1$ channel exhibits a pronounced sensitivity to coupled-channel effects, i.e., the inclusion of $S$–$D$ mixing results in a repulsive $\Lambda_c p$ interaction; its absence leads to a weakly attractive one. Consequently, the spin-averaged correlation function--dominated by the triplet state weight--exhibits repulsive behavior when the $S$–$D$ mixing is present. Furthermore, the source size dependence of the correlation functions is examined, demonstrating that the resulting variations remain experimentally resolvable within the precision of current femtoscopic measurements. A systematic comparison with non-relativistic chiral effective field theory and phenomenological models yields distinct discrepancies in the femtoscopic correlation functions. These findings underscore the capacity of femtoscopy to discriminate between different theoretical descriptions of the $\Lambda_c N$ interaction and provide useful references for upcoming experimental data.
\end{abstract}


\maketitle

\section{INTRODUCTION}
Since the discovery of hyperons in cosmic rays in the early 1950s~\cite{Danysz1953}, hadrons containing strange quarks have attracted extensive interest in both experimental and theoretical studies. Beyond the up, down, and strange quarks, heavier flavor quarks such as charm quarks also exist and can interact with light quarks to form exotic hadrons~\cite{Friday:2025gpj}. Among them, the $\Lambda_c^+$ baryon was the first charmed baryon confirmed in experiment~\cite{Cazzoli:1975et,Knapp:1976qw}, and its quark structure is closely analogous to that of the $\Lambda$ hyperon, with the strange quark replaced by a charm quark~\cite{ParticleDataGroup:2024cfk}. Charm hypernuclei have recently emerged as a subject of growing interest~\cite{Vidana:2019amb,Haidenbauer:2020uci,Wu:2020nin,Liu:2023txn,Yang:2024ats}. In analogy with hypernuclei, whose properties are largely governed by the $\Lambda N$ interaction, the stability of charmed hypernuclei is anticipated to be primarily controlled by the $\Lambda_c N$ interaction. FAIR and J-PARC are expected to produce sufficient charmed particles to generate more charmed hypernuclei~\cite{Riedl:2007sv,Wiedner:2011mf}. Therefore, a detailed understanding of the $\Lambda_c N$ interaction is essential for elucidating the structure and stability of $\Lambda_c$ hypernuclei.

In the experiment, observing charmed hypernuclei would represent a significant milestone and provide valuable insights into their interactions. Notably, the production and study of charmed nuclei have been identified as key physics objectives at major accelerator facilities. At FAIR, the CBM experiment is well suited for reconstructing charmed hadron decays, enabling experimental searches for charmed nuclei, with the charmed deuteron and charmed triton emerging as the most promising candidates based on the $\Lambda_c^{+} \rightarrow p K^{-} \pi^{+}$ decay mode~\cite{Friday:2025gpj}. Meanwhile, the J-PARC facility, benefiting from its high-intensity and high-momentum proton beams, has also been recognized as an ideal facility for studying the production and spectroscopy of charmed nuclei~\cite{Aoki:2021cqa}. 

Given the limited experimental data, theoretical studies of the $\Lambda_c N$ interaction are essential for understanding the properties of charmed nuclei and for guiding future experiments. Early theoretical investigations based on meson-exchange frameworks~\cite{Dover:1977jw,Bando:1981ti} and constituent quark models~\cite{Maeda:2015hxa,Froemel:2004ea} have indicated that the $\Lambda_c N$  interaction is strongly attractive. More recently, the HAL QCD Collaboration has explored the $\Lambda_c N$  interaction through lattice QCD simulations performed at unphysical pion masses~\cite{Miyamoto:2017tjs,Miyamoto:2017ynx}. Later, Haidenbauer and Krein obtained the $\Lambda_c N$ potential at the physical pion mass with chiral effective field theory(ChEFT) by extrapolating the HAL QCD results at unphysical pion masses~\cite{Haidenbauer:2017dua}. They suggested that the  $\Lambda_c N$ interactions could be much less attractive than predicted by the phenomenological potentials mentioned above. 

Recently, femtoscopy--a technique that analyzes momentum correlations between particles emitted in high-energy collisions--has emerged as a powerful alternative for probing the strong interaction. By measuring momentum correlation functions (CFs), femtoscopy has provided valuable insights into rare hadron-hadron scattering processes~\cite{Fabbietti:2020bfg,Liu:2024uxn}. To date, a large variety of hadron-hadron interactions have been extensively studied by the ALICE and STAR collaborations, such as  meson-meson~\cite{ALICE:2017jto,ALICE:2018nnl,ALICE:2020mkb,ALICE:2021ovd,ALICE:2023eyl}, meson-baryon~\cite{ALICE:2019gcn,ALICE:2020wvi,ALICE:2021cpv,ALICE:2022yyh,ALICE:2023wjz,ALICE:2025aur,ALICE:2025flv,ALICE:2025kma}, and (anti)baryon-(anti)baryon~\cite{ALICE:2019buq,STAR:2005rpl,STAR:2014dcy,STAR:2015kha,ALICE:2018ysd,STAR:2018uho,ALICE:2019eol,ALICE:2019igo,ALICE:2019hdt,ALICE:2020ibs,ALICE:2020mfd,ALICE:2021njx,Isshiki:2021bqh,ALICE:2021cyj,ALICE:2022uso,ALICE:2023bny,ALICE:2025plu,ALICE:2025byl} interactions in the light quark (u, d, s) sector with the femtoscopic technique. More recently, femtoscopic measurements involving charm hadrons have demonstrated a great potential of this approach for accessing the charm sector experimentally~\cite{ALICE:2022enj,ALICE:2024bhk}. In particular, femtoscopic measurements of the $p$–$\Lambda_c^+$ system are currently being pursued by the ALICE Collaboration~\cite{Zhang:2025szg}, which is welcome news for extracting the $\Lambda_c N$ interaction. These experimental developments call for reliable theoretical descriptions of the corresponding correlation functions, enabling quantitative interpretations of present and forthcoming data.

In parallel with these experimental efforts, a substantial body of theoretical work has been devoted to the study of hadron–hadron interactions~\cite{Morita:2014kza,Morita:2016auo,Haidenbauer:2018jvl,Morita:2019rph,Ogata:2021mbo,Kamiya:2021hdb,Haidenbauer:2021zvr,Liu:2022nec,Molina:2023jov,Molina:2023oeu,Sarti:2023wlg,Ge:2025put}, especially in the charm sector~\cite{Haidenbauer:2020kwo,Kamiya:2022thy,Liu:2023uly,Liu:2023wfo,Vidana:2023olz,Ikeno:2023ojl,Albaladejo:2023pzq,Torres-Rincon:2023qll,Albaladejo:2023wmv,Shen:2024npc,Liu:2024nac,Liu:2025rci,Liu:2025oar,Xie:2025xew,Liu:2025wwx,Shen:2025jmy,Liu:2025nze}. Among these studies, Ref.~\cite{Haidenbauer:2020kwo} investigated the correlation functions corresponding to the weakly attractive $\Lambda_c N$ interaction predicted by non-relativistic ChEFT. In addition, within a covariant ChEFT, the $YN$ and $YY$  systems with strangeness from $S=-1$ to $S=-4$ have been studied~\cite{Li:2016paq,Li:2016mln,Song:2018qqm,Li:2018tbt,Liu:2020uxi,Song:2021yab,Liu:2022nec}, and this framework has subsequently been extended to the $Y_c N$ system~\cite{Song:2020isu}. The covariant treatment of baryon spinors preserves all relevant symmetries and exhibits improved convergence behavior~\cite{Lu:2021gsb}. The present paper aims to provide quantitative theoretical predictions for the $\Lambda_c N$ correlation function, in anticipation of ongoing measurements by the ALICE Collaboration~\cite{Zhang:2025szg}, based on covariant ChEFT. 

The paper is organized as follows. In Sec.~\ref{Theoretical Framework}, we outline the theoretical framework for correlation functions and introduce the covariant ChEFT approach for the $Y_c N$ system. In Sec.~\ref{results}, we present the fitted results and their extrapolations, and provide predictions for the corresponding $\Lambda_c p$ correlation functions. Finally, a summary is given in the last section.

\section{THEORETICAL FRAMEWORK}\label{Theoretical Framework}

\subsection{ Correlation functions}
The Koonin–Pratt (KP) formalism provides the framework for computing two-hadron momentum correlation functions~\cite{Pratt:1990zq,Bauer:1992ffu,Koonin:1977fh},
\begin{equation}
\begin{aligned}
C\left(\boldsymbol{p}_1, \boldsymbol{p}_2\right) & =\frac{\int \mathrm{d}^4 x_1 \mathrm{~d}^4 x_2 S_1\left(x_1, \boldsymbol{p}_1\right) S_2\left(x_2, \boldsymbol{p}_2\right)\left|\Psi^{(-)}(\boldsymbol{r}, \boldsymbol{k})\right|^2}{\int \mathrm{~d}^4 x_1 \mathrm{~d}^4 x_2 S_1\left(x_1, \boldsymbol{p}_1\right) S_2\left(x_2, \boldsymbol{p}_2\right)} \\
& \simeq \int \mathrm{d} \boldsymbol{r} S_{12}(r)\left|\Psi^{(-)}(\boldsymbol{r}, \boldsymbol{k})\right|^2,
\end{aligned}
\end{equation}
where $S_i\left(x_i, \boldsymbol{p}_i\right)(i=1,2)$ denotes the single-particle source function of hadron $i$ with momentum $\boldsymbol{p}_i$. In the center-of-mass (c.m.) frame, $\Psi^{(-)}(\boldsymbol{r}, \boldsymbol{k})$ is the relative wave function with relative coordinate $\boldsymbol{r} $ and relative momentum $\mathbf{k}=\left(m_2 \mathbf{p}_1-m_1 \mathbf{p}_2\right) /\left(m_1+m_2\right)$, embedding the effects of final-state interactions.  In the present study, we adopt the usual approximation of a static spherical Gaussian source with radius $R$, given by $S_{12}(r)=\exp \left(-r^2 / 4 R^2\right) /(2 \sqrt{\pi} R)^3$.

As $S$-wave interactions predominantly drive the correlations, we consider only the $S$-wave component of the relative wave function to be affected by final-state interactions~\cite{Ohnishi:2016elb},
\begin{equation}\label{PIS}
\Psi_S^{(-)}(\boldsymbol{r}, \boldsymbol{k})=\mathrm{e}^{\mathrm{i} \boldsymbol{k} \cdot \boldsymbol{r}}+\psi_0(r, k)-j_0(k r),
\end{equation}
where $\psi_0$ denotes the $l=0$ scattering wave function affected by the strong interaction and $j_0$ is the $S$-wave component of the non-interacting wave function, represented by a spherical Bessel function. Then, the correlation function becomes:
\begin{equation}
C(k)=1+4 \pi \int d r r^2 S_{12}(r)\left[\left|\psi_0(r,k)\right|^2-\left|j_0(k r)\right|^2\right],
\end{equation}
$\psi_0$ can be determined either by solving the Schroedinger equation for a given potential or via the Lippmann–Schwinger (LS) or Kadyshevsky equation~\cite{HADES:2016dyd,Haidenbauer:2020kwo,ALICE:2021njx}. In covariant ChEFT, the leading-order (LO) four-baryon contact potentials are momentum-dependent (non-local) due to the inclusion of the small component of the Dirac spinor~\cite{Li:2016mln,Ren:2016jna}. It is convenient to first obtain the reaction amplitude $T$ by solving the Kadyshevsky equation, and then derive the corresponding scattering wave function. In the single-channel case, after partial-wave expansion, this reads~\cite{Haftel:1970zz}:
\begin{equation}\label{PIT}
\begin{aligned}
\tilde{\psi}_0(r,k)= &j_0(k r) +\frac{1}{\pi} \int d q q^2 j_0(q r)\\  
&\times \frac{1}{\sqrt{s}-E_1(q)-E_2(q)+i \epsilon} T(q, k ; \sqrt{s}),
\end{aligned}
\end{equation}
where $T(q, k ; \sqrt{s})$ denote the $S$-wave component of the half-off-shell T-matrix , $\tilde{\psi}_0(r,k)=\exp \left(-2 i \delta\right) \psi_0(r,k)$, and $\delta$ is the phase shift. The total energy of
the baryon-baryon system is defined as $\sqrt{s}=E_1(k)+E_2(k)$, with $E_i(k)=\sqrt{k^2+m_i^2}$. The normalization of $\psi_0(r,k)$ is ~\cite{Ohnishi:2016elb},
\begin{equation}
\psi_0(r, k) \stackrel{r \rightarrow \infty}{\longmapsto} \frac{1}{2 \mathrm{i} k r}\left[\mathrm{e}^{\mathrm{i} k r}-\mathrm{e}^{-2 \mathrm{i} \delta} \mathrm{e}^{-\mathrm{i} k r}\right].
\end{equation}

As the measured correlation functions are spin-averaged, the theoretical ones should also be averaged over the total spin of the hadron pair, using the appropriate weights for the spin-singlet and spin-triplet $S$-wave states, namely  $\left|\psi_0\right|^2 \rightarrow 1 / 4\left|\psi_{^1 S_0}\right|^2+3 / 4\left|\psi_{^3 S_1}\right|^2$.

For the $\Lambda_c p$ system, considering
the strong and Coulomb interactions, the relative wave function reads~\cite{Morita:2016auo} 
\begin{equation}\label{coulpi}
\Psi_{S C}^{(-)}(\boldsymbol{r}, \boldsymbol{k})=\phi^C(\boldsymbol{r}, \boldsymbol{k})+\psi_0^{S C}(r, k)-\phi_0^C(k r),
\end{equation}
where $\phi^C(\boldsymbol{r}, \boldsymbol{k})$ and $\phi_0^C(k r)$ denote the full Coulomb wave function and its $S$-wave component, $\psi_0^{S C}(r, k)$ is the scattering wave function incorporating both strong and Coulomb interactions, obtained as in Eq.~(\ref{PIT}). In the present study, the Coulomb interaction is treated in momentum space via the Vincent–Phatak method~\cite{Haidenbauer:2020kwo,Vincent:1974zz,Holzenkamp:1989tq}. By substituting the relative wave function~(\ref{coulpi}) into the KP formula, the correlation function becomes
\begin{equation}
\begin{aligned}
C(k) = & \int \mathrm{d} \boldsymbol{r} S_{12}(r)\left|\phi^C(\boldsymbol{r}, \boldsymbol{k})\right|^2 \\
&+4 \pi\int_0^{\infty}  r^2 \mathrm{~d} r S_{12}(r)\left[\left|\psi_0^{S C}(r, k)\right|^2-\left|\phi_0^C(k r)\right|^2\right].
\end{aligned}
\end{equation}

\subsection{ $Y_{c} N$ interactions in the leading-order covariant ChEFT}
In this section, we briefly outline the $Y_{c} N$ potential. It is worth noting that we have corrected an inconsistency in Ref.~\cite{Song:2020isu}, where the low-energy constants (LECs) for the $^1S_0$ and $^3S_1$ terms were mixed. At leading order, the $Y_{c} N$ potentials consist of contributions from nonderivative four-baryon contact terms (CTs) and one-meson exchanges (OMEs)~\cite{Li:2016mln,Song:2020isu}. For the $^1S_0$ and $^3S_1-^3D_1$ partial waves considered here, one obtains: 
\begin{widetext}
\begin{equation}
  V_{^1S_0}^{Y_{c} N}=\xi_{Y_{c} N}\left[C_{1S0}^{Y_{c} N}\left(1+R^{Y_{c}}_{p^\prime} R^{Y_{c}}_pR^N_{p^\prime} R^N_p\right)+\hat{C}_{1S0}^{Y_{c} N} \left(R^{Y_{c}}_{p^\prime}R^N_{p^\prime}+R^{Y_{c}}_pR^N_p\right)\right],
\end{equation}
\begin{equation}
  V_{^3S_1}^{Y_{c} N}=\xi_{Y_{c} N}\left[\frac{1}{9}C_{3S1}^{Y_{c} N}\left(9+R^{Y_{c}}_{p^\prime}R^{Y_{c}}_pR^N_{p^\prime}R^N_p\right)+\frac{1}{3}\hat{C}_{3S1}^{Y_{c} N}\left(R^{Y_{c}}_{p^\prime}R^N_{p^\prime}+R^{Y_{c}}_pR^N_p\right)\right], 
\end{equation}
\begin{equation}
   V_{^3D_1}^{Y_{c} N}=\xi_{Y_{c} N}\left[\frac{8}{9}C_{3S1}^{Y_{c} N}  R^{Y_{c}}_{p^\prime}R^{Y_{c}}_pR^N_{p^\prime}R^N_p\right],   
\end{equation}
\begin{equation}
   V_{^3SD_1}^{Y_{c} N}=\xi_{Y_{c} N}\left[\frac{2\sqrt{2}}{9}C_{3S1}^{Y_{c} N} R^{Y_{c}}_{p^\prime}R^{Y_{c}}_pR^N_{p^\prime}R^N_p+\frac{2\sqrt{2}}{3}\hat{C}_{3S1}^{Y_{c} N}R^{Y_{c}}_pR^N_p\right],  
\end{equation}
\begin{equation}
    V_{^3DS_1}^{Y_{c} N}=\xi_{Y_{c} N}\left[\frac{2\sqrt{2}}{9}C_{3S1}^{Y_{c} N} R^{Y_{c}}_{p^\prime}R^{Y_{c}}_pR^N_{p^\prime}R^N_p+\frac{2\sqrt{2}}{3}\hat{C}_{3S1}^{Y_{c} N}R^{Y_{c}}_{p^\prime}R^N_{p^\prime}\right],  
\end{equation}
where
\begin{equation}
\xi_{Y_{c} N}=4 \pi \frac{\sqrt{\left(E_{p^{\prime}}^{Y_{c}}+M_{Y_{c}}\right)\left(E_p^{Y_{c}}+M_{Y_{c}}\right)\left(E_{p^{\prime}}^N+M_N\right)\left(E_p^N+M_N\right)}}{4 M_N M_{Y_{c}}} \quad \text { and } \quad R_{p\left(p^{\prime}\right)}^{Y_{c}, N}=\frac{p\left(p^{\prime}\right)}{E_{p\left(p^{\prime}\right)}^{Y_{c}, N}+M_{Y_{c}, N}}.
\end{equation}
\end{widetext}

The OME potentials have the following form,
\begin{widetext}
\begin{equation}
V^{Y_{c} N \rightarrow Y_{c} N}_{\mathrm{OME}}=N \frac{\left(\bar{u}_{Y_{c}}(p^\prime) \gamma^\mu \gamma_5 q_\mu u_{Y_{c}}(p)\right)\left(\bar{u}_N(-p^\prime) \gamma^v \gamma_5 q_v u_N(p)\right)}{q^2-m^2} \mathcal{I}_{Y_{c} N \rightarrow Y_{c} N},
\end{equation}
\end{widetext}
where $Y_{c}=\Lambda_{c},\Sigma_{c}$, $q=\left(E_{p^{\prime}}-E_p, \boldsymbol{p^{\prime}}-\boldsymbol{p}\right)$ represents the four-momentum transferred, and $m$ is the mass of the exchanged pseudoscalar
meson. $N$ and $\mathcal{I}_{Y_{c} N \rightarrow Y_{c} N}$ are the coupling constants and isospin factors, respectively, as listed in Refs.~\cite{Polinder:2006zh,Li:2016mln,Haidenbauer:2020kwo}. The ${C}_{1S0}$, $\hat{C}_{1S0}$,
${C}_{3S1}$, and $\hat{C}_{3S1}$ are the LECs in the CT potentials.  For the $Y_c N$ interactions, they are constrained by fitting to the lattice QCD simulations from the HAL QCD Collaboration~\cite{Miyamoto:2017tjs}.

In addition, to avoid ultraviolet divergence in numerical evaluations,
the $Y_c N$ potentials are regularized with an exponential form factor,
\begin{equation}
f_{\Lambda_F}\left(\boldsymbol{p}, \boldsymbol{p}^{\prime}\right)=\exp \left[-\left(\frac{\boldsymbol{p}}{\Lambda_F}\right)^{2 n}-\left(\frac{\boldsymbol{p}^{\prime}}{\Lambda_F}\right)^{2 n}\right],
\end{equation}
where $n = 2$ and $\Lambda_F$ is the cutoff momentum. We consider cutoff values from 500 to 600 MeV in the present work. The resulting uncertainties indicate the cutoff dependence and thus provide a lower bound on theoretical uncertainties.

\section{RESULTS AND DISCUSSION}\label{results}

\subsection{Fitting Results of the $Y_c N$ interactions}

In this work, we refitted the $Y_c N$ interactions because the revised potential formulation required corresponding updates to the LECs. As mentioned above, the LECs of $Y_c N$ interactions are determined by fitting to the lattice QCD simulations from the HAL QCD Collaboration~\cite{Miyamoto:2017tjs}. We used the $\Lambda_c N$ $S$-wave phase shifts
for $m_\pi$ = 410 and 570 MeV with the center-of-mass energy $E$ up to 30 MeV.
Following Refs.~\cite{Haidenbauer:2017dua,Song:2020isu}, we adopted an effective CT potential that includes only the $\Lambda_cN \rightarrow \Lambda_cN$  channel, assuming that the $\Sigma_cN$ contributions can be absorbed implicitly. Thus, only four LECs were needed in this work: $C_{1S0}$,$\hat C_{1S0}$,$C_{3S1}$,$\hat C_{3S1}$. As demonstrated in Ref.~\cite{Song:2020isu}, the covariant ChEFT predictions for the spin-triplet channel exhibit a pronounced sensitivity to the treatment of coupled-channel effects. Motivated by this observation, we considered two fitting strategies for the CT term: one that explicitly includes the $S-D$ mixing and another that does not. The fitted LECs are shown in
Table~\ref{lecs}.

\begin{table}[H]
\centering
\caption{Low-energy constants (in units of $10^4$ GeV$^{-2}$) obtained for two cutoffs of $\Lambda_F=500$ and 600 MeV in the covariant ChEFT, using fitting schemes with and without $S-D$ mixing. These LECs are determined by fitting the lattice QCD simulations from the HAL QCD Collaboration~\cite{Miyamoto:2017tjs}.}
\setlength{\tabcolsep}{1.1pt}
\begin{tabular}{lcccccccc}
\hline \hline$m_\pi$ & $\Lambda_F$ & $C_{1S0}$ & $\hat C_{1S0}$ & $C^{w/. S-D}_{3S1}$ &  $\hat C^{w/. S-D}_{3S1}$ &  $C^{w/o. S-D}_{3S1}$ & $\hat C^{w/o. S-D}_{3S1}$  \\
\hline
410 & 500 & $-$0.0150 & 1.4069 & 0.2306 & 16.8820 & $-$0.0162 & 4.0940 \\
    & 600 & $-$0.0124 & 1.6158 & 0.0020 & $-$2.7921 & $-$0.0141 & 4.5090 \\
\hline
570 & 500 & $-$0.0079 & 0.5596 & $-$0.0042 & 2.3141 & $-$0.0093 & 1.7227 \\
    & 600 & $-$0.0074 & 0.6204 & 0.1605 & 10.2650 & $-$0.0088 & 1.8317 \\
\hline \hline
\end{tabular}
\label{lecs}
\end{table}

\begin{figure}[H]
    \centering
    \includegraphics[scale=0.35]{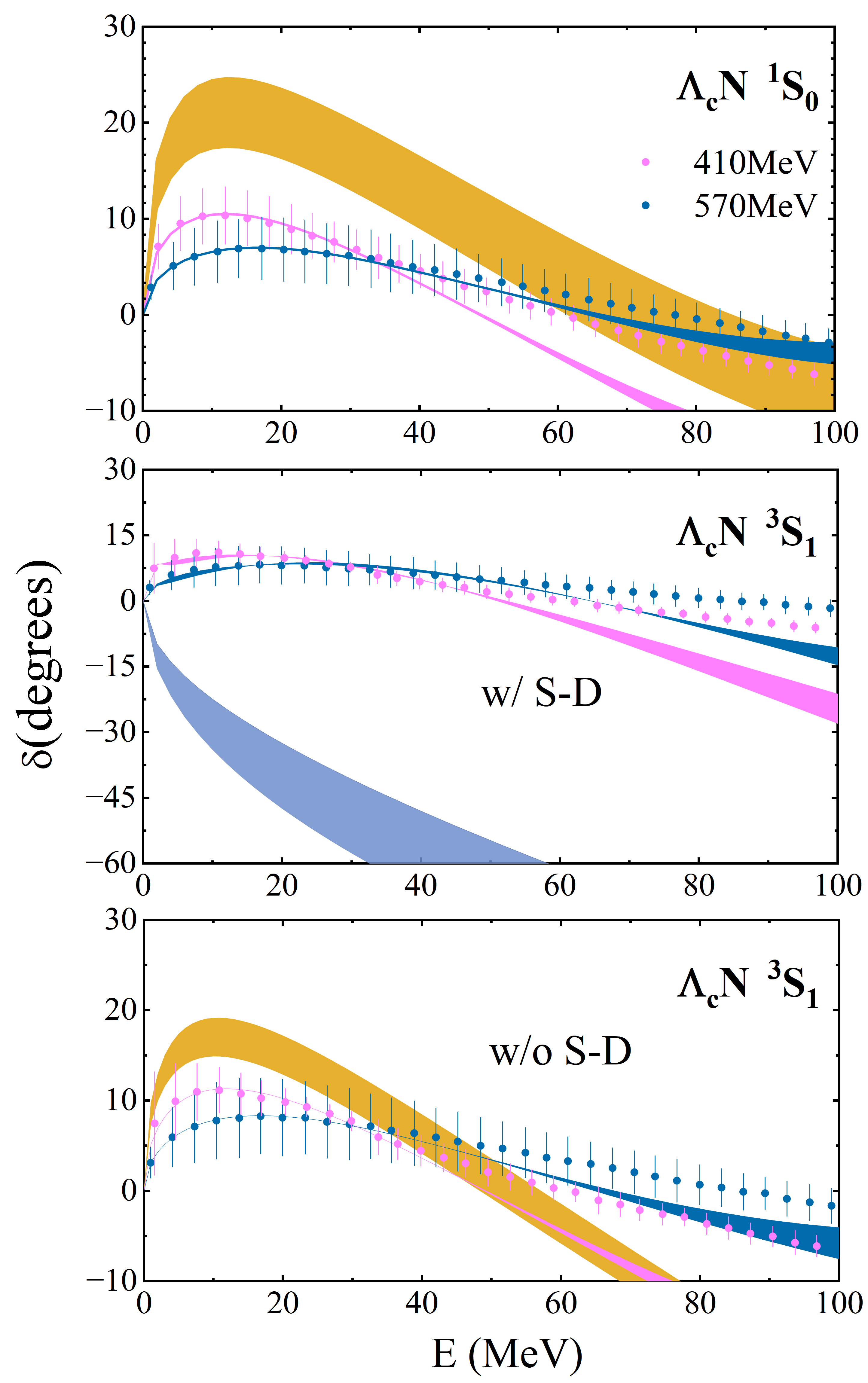}
    \caption{$\Lambda_c N$ $S$-wave phase shifts of the lattice QCD simulations in
comparison with the ChEFT fits. Magenta and dark-blue dots correspond to the LQCD data~\cite{Miyamoto:2017tjs} at $m_{\pi}=410$ MeV and 570 MeV, respectively. Magenta and dark blue bands show the ChEFT fits with $\Lambda_F$ ranging from 500 to 600 MeV. Yellow and light-blue bands denote ChEFT predictions at $m_{\pi}=138$ MeV. The $\Lambda_c N$ interaction is obtained by fitting to the lattice QCD data only up to $E=30$ MeV. }
    \label{Phaseshifts}
\end{figure}

The fitted $S$-wave $\Lambda_c N$ phase shifts are shown in Fig.~\ref{Phaseshifts}.  The covariant ChEFT results agree well with the lattice QCD data. For the $^1S_0$ partial wave, the extrapolated  $\Lambda_c N$ interaction at the physical pion mass is moderately attractive. In the case of the $^3S_1$ channel, displayed in the middle panel of Fig.~\ref{Phaseshifts}, the covariant ChEFT phase shifts are in fair
agreement with the lattice QCD data for energies up to
50 MeV. The discrepancy then becomes larger as the energy
increases. The interaction becomes increasingly repulsive as the energy increases when extrapolated to the physical pion mass.  To clarify the origin of this behavior, we set $V_{{ }^3 SD_1}$
and  $V_{{ }^3 DS_1}$ in the CT potential to zero and redid the fits. The
resulting phase shifts are shown in the bottom panel of Fig.~\ref{Phaseshifts}. When extrapolated to the physical point, an attractive interaction is obtained, which is similar not only to the
$^1S_0$ interaction, but also to the $^3S_1$ interaction of the non-relativistic ChEFT. As a result, we conclude that the predicted $^3S_1$ interaction depends strongly on how the coupled channel $S-D$ mixing is treated. The results are consistent with those of Ref.~\cite{Song:2020isu}; a more detailed analysis can be found therein.

\begin{figure}[H]
    \centering    \includegraphics[scale=0.37]{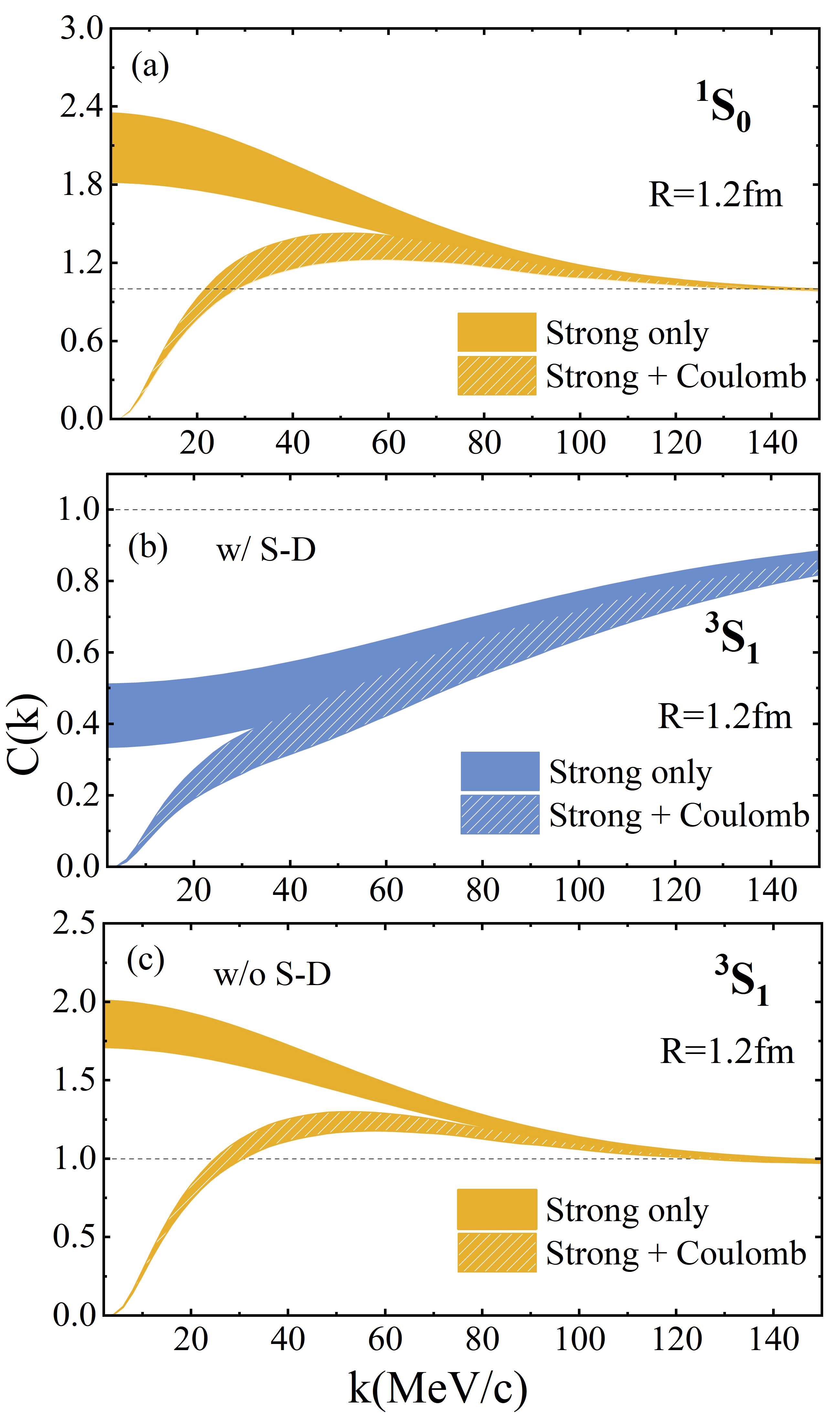}
    \caption{Effect of the Coulomb interaction on the $\Lambda_c   p$ correlation function in the $^1S_0$ and $^3S_1$ partial waves. (a) The yellow (yellow shaded) band shows the results without (with) the Coulomb interaction.
(b) Similar to panel (a), the $^{3}S_{1}$ results with $S-D$ mixing are shown.
(c) Similar to panel (a), the $^{3}S_{1}$ results without $S-D$ mixing are shown. }
\label{1S0 3S1 CF}
\end{figure}

In Fig~\ref{1S0 3S1 CF}, we show the $\Lambda_c p$ correlation functions for the $^1S_0$ and $^3S_1$ partial waves, selectively for the source radius $R=1.2$ fm. In the case of strong interactions only, the $^1S_0$ and $^3S_1$ without $S-D$ mixing exhibit attractive behavior, whereas the $^3S_1$ with $S-D$ mixing shows repulsion. These trends are consistent with the corresponding phase shift results shown in
Fig~\ref{Phaseshifts}.  Once the Coulomb interaction is taken into account, the repulsive Coulomb force in the $\Lambda_c p$ system leads to a pronounced suppression of the correlation function at small relative momenta. The behavior is well established and discussed in Ref.~\cite{ALICE:2018ysd}.  Notably, the $\Lambda_c p$  attraction is substantially weaker than that of the $pp$ system. As a consequence, the Coulomb repulsion dominates at higher momenta, leading to enhanced depletion of the correlation function and a shift of its maximum to higher momenta. This competition results in a substantial suppression of the strong interaction contribution.

\begin{figure}[hbt]
    \centering    \includegraphics[scale=0.37]{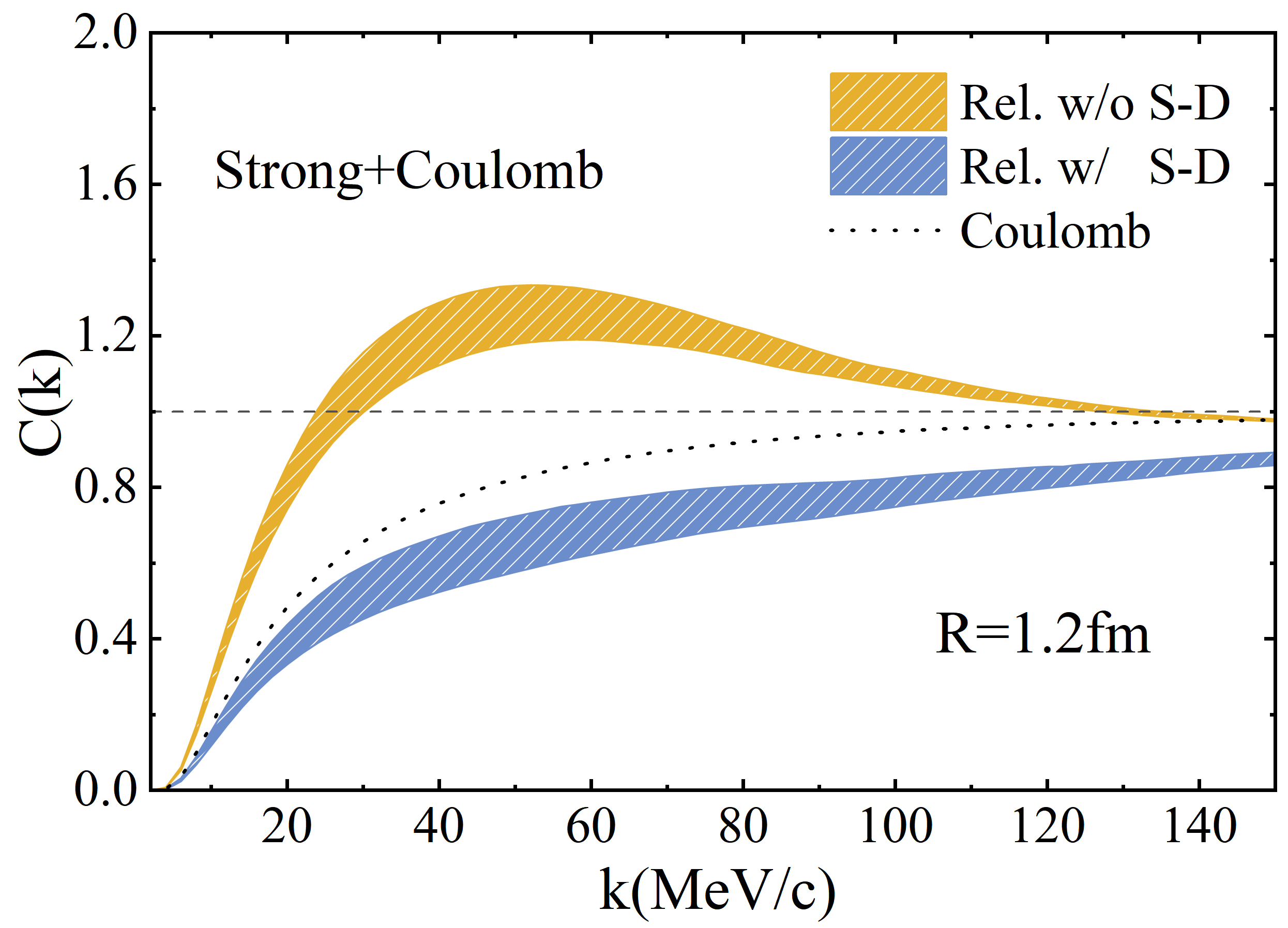}
    \caption{Spin-averaged $\Lambda_c p$ correlation functions for $R=1.2$ fm with the Coulomb interaction. The covariant ChEFT with $S-D$ mixing (blue) and without $S-D$ mixing (yellow) are shown. The dotted curve denotes the result obtained with the pure Coulomb interaction.}
    \label{Spin average CF}
\end{figure}

Next, to better match the experimental situation, we consider the spin-averaged $\Lambda_c p$ correlation functions. In Fig.~\ref{Spin average CF}, we present the results predicted by the covariant ChEFT. The yellow band corresponds to the result obtained without $S$–$D$ mixing, while the blue band represents the results obtained with $S$–$D$ mixing. For comparison, the dotted line indicates the result obtained with the pure Coulomb interaction. The spin-averaged $\Lambda_c p$ correlation functions exhibit clearly different behaviors depending on the presence or absence of $S$–$D$ mixing. The result without $S$–$D$ mixing becomes more attractive compared with the pure Coulomb correlation, whereas the result with $S$–$D$ mixing turns more repulsive. Although the $^1S_0$ channel becomes more attractive than the $^3S_1$ channel once $S$–$D$ mixing is included (see Fig.~\ref{1S0 3S1 CF}), the resulting spin-averaged interaction remains repulsive, since the $^3S_1$ contribution enters with a statistical weight three times larger than that of the $^1S_0$ channel. 


\begin{figure}[H]
    \centering    \includegraphics[scale=0.37]{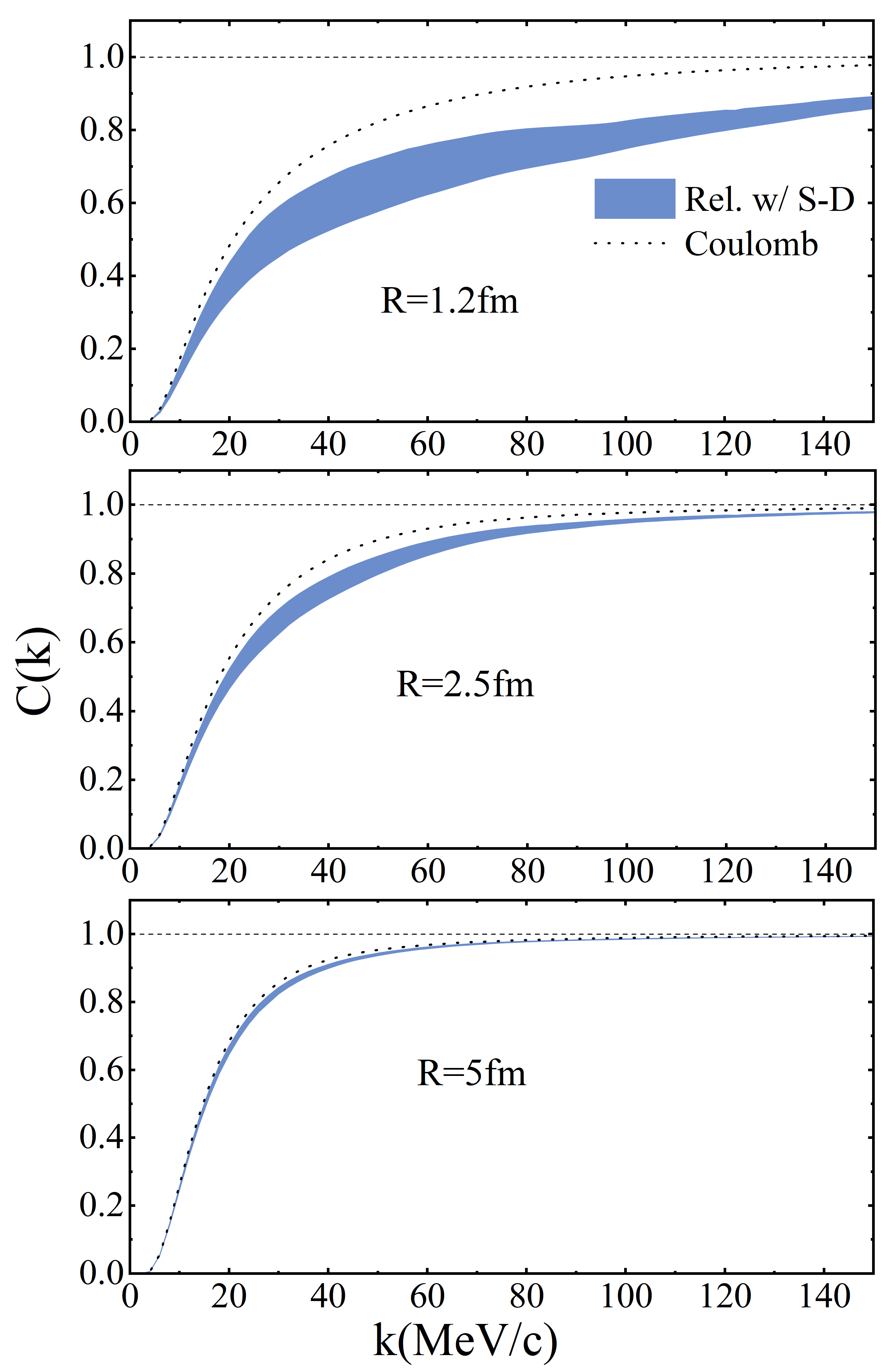}
    \caption{The spin-averaged $\Lambda_c p$ correlation functions that include the Coulomb interaction are presented for three different source radii $R$. Predictions from the covariant ChEFT with $S$–$D$ mixing are presented. For reference, the results obtained with the pure Coulomb correlation are displayed as a dotted line.}
    \label{CF-R}
\end{figure}

Fig.~\ref{CF-R} displays the momentum dependence of the spin-averaged $\Lambda_c p$ correlation functions for different source radii: $R = 1.2, 2.5$, and $5.0$ fm. As the source radius increases, the correlation functions predicted by covariant ChEFT with $S$–$D$ mixing approach the pure Coulomb result (represented by the dotted lines). For the smallest radius, $R = 1.2$ fm, the correlation function exhibits a noticeable deviation from the pure Coulomb baseline, indicating that the strong interaction—specifically the repulsive nature of the triplet channel—remains influential at short distances. However, as $R$ increases to $5.0$ fm, the strong-interaction contribution is significantly diluted. Furthermore, the figure shows a clear reduction in cutoff sensitivity (as indicated by the width of the blue bands) as the source radius increases. For $R = 1.2$ fm, the theoretical uncertainty originating from the LECs and the regulator is more pronounced, reflecting the sensitivity of small-source femtoscopy to the short-range details of the covariant ChEFT potential. In contrast, for $R = 5.0$ fm, the band narrows substantially, demonstrating that the long-distance behavior of the correlation function is robust and less dependent on the specific choice of the high-momentum regulator. These results confirm that while large source radii provide a clean environment to study Coulomb effects, small-source measurements are indispensable for extracting the dynamical details of the $\Lambda_c N$ strong interaction.


\begin{figure}[hbt]
    \centering    \includegraphics[scale=0.30]{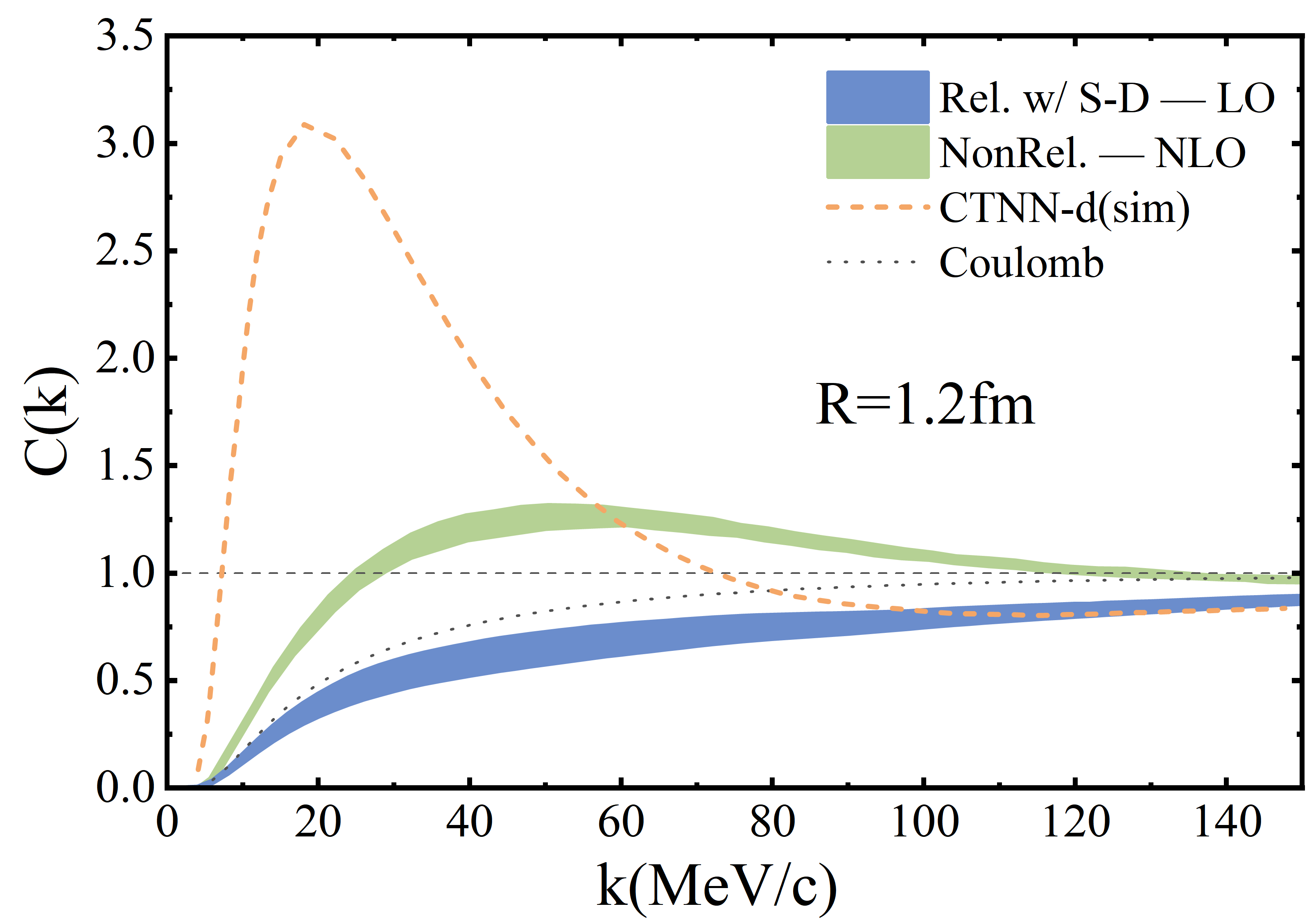}
    \caption{Spin-averaged $\Lambda_c p$ correlation functions including the Coulomb interaction for $R = 1.2$ fm. Results obtained from covariant ChEFT (blue band), non-relativistic ChEFT (green band)~\cite{Haidenbauer:2020kwo}, the simulation of the CTNN-d model (dashed line)~\cite{Haidenbauer:2020kwo}, and the pure Coulomb interaction (dotted line) are shown.}
    \label{Lambdacp-different methods}
\end{figure}

In Fig.~\ref{Lambdacp-different methods}, the spin-averaged $\Lambda_c p$ correlation functions including the Coulomb interaction predicted by covariant ChEFT are compared with the corresponding results from non-relativistic ChEFT~\cite{Haidenbauer:2017dua} and the phenomenological CTNN-d potential~\cite{Maeda:2015hxa}. The latter two results are taken from Ref.~\cite{Haidenbauer:2020kwo}. For a source radius of $R = 1.2$ fm, the different potentials yield markedly different predictions for the correlation function. In general, a more attractive interaction leads to an enhanced correlation function. 
The covariant ChEFT interaction, including $S$–$D$ mixing, yields a smooth, moderate momentum dependence of the correlation function. After spin averaging, the resulting correlation function suggests a weak repulsion. The cutoff dependence is reflected in the width of the blue band, whereas the overall behavior remains robust. The non-relativistic ChEFT interaction produces a more pronounced enhancement at low and intermediate momenta compared to the covariant ChEFT. The correlation function exceeds unity in the region of $k \sim 30$–$100$ MeV/$c$, suggesting a shallow attractive component in this framework. In contrast, the CTNN-d interaction is characterized by a much stronger attraction and supports bound states in both $S$-wave channels with binding energies of the order of the deuteron. Consequently, the predicted correlation function shows a pronounced low-momentum enhancement, featuring a sharp peak around $k \sim 10$–$30$ MeV/$c$. This qualitative behavior is markedly different from that obtained within covariant and non-relativistic ChEFT.

\section{CONCLUSION}
In this work, we have performed a systematic study of the $\Lambda_c p$ momentum correlation functions within the framework of covariant chiral effective field theory at leading order. Our analysis reveals that the $\Lambda_c N$ interaction is moderately attractive in the spin-singlet ${}^1S_0$ channel. The ${}^3S_1$ channel is highly sensitive to the treatment of $S$–$D$ mixing.  While the inclusion of $S$–$D$ mixing results in a repulsive $\Lambda_c p$ interaction, its absence leads to a weakly attractive interaction. Consequently, the spin-averaged correlation function—dominated by the triplet state weight—exhibits a repulsive behavior when $S$–$D$ mixing is present. The correlation functions exhibit distinct behaviors across different source radii ($R = 1.2, 2.5, 5.0$ fm). While the strong interaction signal is most pronounced for small $R$, the correlation functions approach the pure Coulomb limit as $ R$ increases. This confirms that small-source femtoscopy at LHC/RHIC is a viable tool for extracting $\Lambda_c N$ dynamics. A comparative study with non-relativistic ChEFT and the phenomenological CTNN-d potential reveals significant discrepancies. Unlike bound-state models that predict sharp low-momentum peaks, the covariant ChEFT suggests a more moderate, repulsive interaction. Such differences demonstrate that experimental femtoscopic data can effectively discriminate between different theoretical descriptions of the charmed baryon-nucleon force.

In summary, we have provided theoretical predictions for the $\Lambda_c p$ interaction. The identified sensitivity to coupled-channel effects and covariant corrections offers a new perspective on the charmed-hadron interactions, serving as a useful guide for upcoming femtoscopy programs.

\emph{Acknowledgments.} This work is partly supported by the National Key R\&D Program of China under Grant No. 2023YFA1606703 and the National Science Foundation of China under Grants No. W2543006 and No. 12435007. Zhi-Wei Liu acknowledges support from the National Natural Science Foundation of China under Grant No.12405133, No.12347180, China Postdoctoral Science Foundation under Grant No.2023M740189, and the Postdoctoral Fellowship Program of CPSF under Grant No.GZC20233381. Ru-You Zheng acknowledges support from the China Scholarship Council scholarship.

\bibliography{LambdacN}


\clearpage
\end{document}